\documentclass[aps,prb,showpacs,twocolumn,groupedaddress]{revtex4}
\usepackage{graphicx}
\newcommand{\nn}{\nonumber}
\newcommand{\be}{\begin{equation}}
\newcommand{\ee}{\end{equation}}
\newcommand{\bea}{\begin{eqnarray}}
\newcommand{\eea}{\end{eqnarray}}
\newcommand{\h}{Hamiltonian }
\renewcommand{\o}{\omega}
\newcommand{\hb}{\hbar}
\newcommand{\la}{\langle}
\newcommand{\ra}{\rangle}
\newcommand{\half}{\frac{1}{2}\,}

\newcommand{\etal}{{\em{et}}\,{\em{al}}.\,}
\newcommand{\nb}{\bar{n}_j}
\begin{document}

\title{Several Small Josephson Junctions in a 
Resonant Cavity:\\ Deviation from the Dicke Model}

\author{W.~A.~Al-Saidi}
\email{Al-Saidi.1@osu.edu}
\author{ D.~ Stroud}
\email{Stroud@mps.ohio-state.edu}
\affiliation{Department of Physics,
The Ohio State University, Columbus, Ohio 43210}

\date{\today}

\begin{abstract}

We have studied quantum-mechanically a system of several small
identical Josephson junctions in a lossless single-mode cavity for
different initial states, under conditions such that the system is at
resonance.  This system is analogous to a collection of identical
atoms in a cavity, which is described under appropriate conditions by
the Dicke model.  We find that our system can be well approximated by
a reduced Hamiltonian consisting of two levels per junction.  The
reduced Hamiltonian is similar to the Dicke Hamiltonian, but contains
an additional term resembling a dipole-dipole interaction between the
junctions.  This extra term arises when states outside the degenerate
group are included via degenerate second-order (L\"{o}wdin)
perturbation theory.  As in the Dicke model, we find that, when N
junctions are present in the cavity, the oscillation frequency due to
the junction-cavity interaction is enhanced by $\sqrt{N}$. The
corresponding decrease in the Rabi oscillation period may cause it to
be smaller than the decoherence time due to dissipation, making these
oscillations observable.  Finally, we find that the frequency
enhancement survives even if the junctions differ slightly from one
another, as expected in a realistic system.

\end{abstract}


\maketitle

\section{Introduction}

In a previous paper, \cite{alsaidi} we numerically studied a small
Josephson junction capacitively coupled to the single mode of a
resonant cavity.  We found that the quantum states of the combined
system are strongly entangled only at certain special values of the
gate voltage across the junction, corresponding to a one- or
two-photon absorption or emission process by the junction Cooper
pairs.  We also showed that at a gate voltage corresponding to the
one-photon process, our results, obtained from exact diagonalization,
agree well with those obtained from a reduced Jaynes-Cummings
Hamiltonian\cite{jaynes} which includes only two junction states and
two photon states.

In the present paper, we extend this work to several similar junctions
placed in a resonant cavity within a radiation wavelength.  
Each
junction is assumed to be voltage biased by an independent voltage
source, but all bias voltages are assumed to be equal.
As in the single-junction case,
we solve for the eigenstates of this coupled system by direct
diagonalization, and compare the results to those obtained from a
reduced Hamiltonian which includes only two energy levels per
junction.  We also compare the interaction to that in a system
first discussed by Dicke, \cite{Dicke} consisting of an assembly of
identical two-level atoms in a single-mode cavity.

Before proceeding further, we note some distinctions between the
system considered here and another system which has recently been
studied experimentally. \cite{barbara}  These experiments involve a two
dimensional array of junctions, such that each junction is coupled to
a single-mode resonant cavity.  For suitable junction and cavity
parameters, the cavity interaction causes the array to phase-lock
and to radiate coherently into the cavity.  The models used to
describe such arrays\cite{filatrella,cawthorne,almaas,harbaugh}
typically involve a semiclassical approximation, appropriate for large
numbers of photons in the cavity.  By contrast, our system
involves only a few photons, and needs to be studied quantum
mechanically.  Another distinction is that, in  Ref.~\onlinecite{barbara}, the junctions are biased onto either the resistive or
the superconducting branch of the I-V characteristic, whereas in the
present model, only the latter branch is considered. 

The present work may also be relevant to quantum
computation. \cite{MSS}  Josephson devices have been proposed as
possible quantum bits (qubits) in computing systems, in which quantum
logic operations could be performed by controlling gate voltages or
magnetic
fields. \cite{MSS,nakamura1,nakamura2,shnirman,mooij,feigelman}  Thus
far, two alternative realizations of qubits have been proposed for
Josephson junctions: a charge qubit and a flux qubit.  In the first
case, the charging energy of the junction is much larger than the
Josephson energy, so that the charge on one of the superconducting
islands is nearly a good quantum number.  Qubits of this type have been
studied experimentally by Nakamura \etal . \cite{nakamura1,nakamura2}
In the second case, studied experimentally by Mooij \etal \cite{mooij}
and Feigel'man \etal , \cite{feigelman} the Josephson energy
predominates, and the phase difference across the junction is well
defined.  In the present paper, the system is constructed from
junctions in which the charging and Josephson energies are comparable
in magnitude; in other respects, it somewhat resembles that studied in
Ref.~\onlinecite{shnirman}.

The findings of this paper may also be relevant for studying ``quantum
leakage''\cite{fazio} in Josephson-junction-based multi-qubits.
Quantum leakage occurs when the computational space of the physical
system is a subspace of a larger Hilbert space.  Such leaking is an
intrinsic source of decoherence, even in the absence of dissipation,
and imposes a time limit beyond which the system can no longer perform
quantum computations.  Thus, comparing the results of numerically
diagonalizing the full Hamiltonian and  the reduced Hamiltonian
provides a means of studying this decoherence. Such a comparison was
done in Ref.~\onlinecite{fazio}, and is also done here for a different
Hamiltonian and a different parameter region.

Finally, because of the analogy between our system and a group of
two-level atoms in a single-mode cavity, the present work may be
relevant to the field of cavity quantum
electrodynamics. \cite{raimond} Specifically, using the system
considered here, it may be possible to ``engineer'' a so-called
Einstein-Podolsky-Rosen or Greenberger-Horne-Zeilinger state comprised
of Josephson junctions.  It may also be feasible to perform a
``quantum nondemolition'' measurement on a one-photon resonator much
as is done using a two-level atom in a cavity. \cite{scully} We hope
to develop these ideas in a future publication.

The remainder of this paper is organized as follows.  In Section II,
we introduce our model Hamiltonian and its physical realization; we
also outline the method used to solve for its eigenstates.  Section
III contains our principal results.  We describe the reduced
Hamiltonian, and we compare the time evolution of the system as
predicted by the full Hamiltonian and by the two-level approximation.
In the final section, the main points of the paper are summarized, and
some of the physical parameters entering the calculations are estimated.

\section{Model Hamiltonian and its Diagonalization}

\subsection{Hamiltonian}

We consider a group of underdamped Josephson junctions in a lossless
electromagnetic cavity which can support a single 
photon mode of angular frequency $\o$.  The system is assumed to be
described by the following model Hamiltonian:
\begin{equation}
{\mathcal{H}} = {\mathcal{H}}_{\mathrm{photon}} +\sum_j
{\mathcal{H}}_{\mathrm{J}}(j) \label{eq:hamtotal}.
\end{equation}
Here ${\mathcal{H}}_{\mathrm{photon}}$ is the \h of the cavity mode,
which we express in the form $ {\mathcal{H}}_{\mathrm{photon}} =
\hbar\,\omega\left(a^\dag a + {1}/{2}\right), $ where $a^\dag$ and $a$
are the photon creation and annihilation operators with commutation
relation $[a,a^{\dagger}]=1$.  ${\mathcal{H}}_{\mathrm{J}}(j)$ is the
\h of the $j$th junction, which implicitly contains the coupling to the
cavity; it can be written as   
\be
{\mathcal{H}}_{\mathrm{J}}(j) = \frac{1}{2}\,U\,(n_j - \bar{n}_j)^2
-J\cos\gamma_j, \label{eq:hamJG} 
\ee 
where the first and second terms
represent the charging and Josephson energies of a junction. $U =
4e^2/C$ is the junction charging energy, $E_J =\hbar I_c/(2e)$ is the
Josephson coupling energy, $C$ is the effective junction capacitance,
and $I_c$ the junction critical current.
$n_j$ is an operator which represents the difference between the
number of Cooper pairs on the two grains comprising the junction, and
$\nb$ is related to the gate voltage applied across the junction. 
Although $\nb$ could, in principle, be controlled separately for each
junction, we assume that all the $\nb$'s are equal.  The last
parameter, $\gamma_j$, is the gauge invariant phase of the junction,
as defined below.  We assume that all the junctions have the same
$J$ and $U$, and also that $J$ and $U$ are comparable in magnitude.
Furthermore, we neglect the current through the shunt resistance.  
Finally, we consider only
junctions such that both $J$ and $U$ are small compared to the 
superconducting gap $\Delta$, which is thus the largest 
energy scale in the system.
 
The interaction between the junctions and the electromagnetic
field is contained in the gauge-invariant phase difference 
\be
\gamma_j = 
        \phi_j - ({2\pi}/{\Phi_0})\int_j {\bf A}({\bf x}, t) \cdot d{\bf \ell}.
\ee
Here $\phi_j$ is the phase difference across the junction in a
particular gauge; it satisfies the commutation relation
$[n_k, \phi_j]= - i\,\delta_{ j\,k}$.  
$\Phi_0=hc/(2 e)$ is the flux quantum, ${\bf A}({\bf x},t)$ 
is the vector potential, and the line integral
is taken across the j${th}$ junction.  We consider only a 
vector potential arising from the electromagnetic field of the 
cavity normal mode.  We also
assume that the junctions are all located within a distance small
compared to the radiation wavelength, so that
${\bf A}({\bf x})$ is approximately uniform throughout the region of the
junctions.  If the junctions all have the same orientation, this
condition implies that each junction has the same coupling to the cavity
electromagnetic field. The generalization to nonuniform fields
is straightforward, however.  

In Gaussian units, and in the Coulomb gauge (${\bf \nabla}~\cdot~
{\bf A} ~= ~0$), the vector potential takes the form
$
{\bf A} = \sqrt{{h\,c^2}/{\omega\, V}}\left( a + a^\dag\right){\hat \epsilon},
$
where ${\hat \epsilon}$ is the unit polarization vector 
of the cavity mode, and $V$ is 
the volume of the cavity.  Thus we can define the coupling $g$ such that
$({2\pi}/{\Phi_0})\int {\bf A} \cdot d{\bf \ell}= ({g}/{\sqrt{2}})(a +
a^\dag),$ or 
\be
g = 4\,e\,\sqrt{\frac{\pi}{\hbar\,\omega\, V}}\, {\hat
\epsilon}\cdot {\vec \ell}, \label{eq:g}
\ee
where $\ell$ is the thickness  of the insulating layer in a junction.
With the change of variables 
$
p = i\sqrt{\hbar\,\omega/{2}}\left(a^\dag - a\right); 
q = \sqrt{{\hbar}/{(2\,\omega)}}\left(a^\dag + a\right),
$
$\gamma_j$ can be rewritten
\be
\gamma_j = \phi_j - g\,\sqrt{{\omega}/{\hbar}}\,q,
\ee
where $p$ and $q$ satisfy the canonical commutation relation
$[p, q] = -i\,\hbar$. 

Finally, we decompose  the \h (\ref{eq:hamtotal}) as follows:
\be
{\mathcal{H}} = {\mathcal{H}}_{\mathrm{photon}} 
+\sum_j {\mathcal{H}}_{J}^{0}(j) +{\mathcal{H}}_{\mathrm{int}},
\ee
where 
\be
{\mathcal{H}}_{J}^{0}(j) = \frac{1}{2}\,U\,(n_j - \bar{n}_j)^2 -J\cos\phi_j,
\label{eq:hJunc}
\ee
and 
\be
{\mathcal{H}}_{\mathrm{int}}= -J\, \sum_j (\cos\gamma_j - \cos\phi_j).
\label{eq:hint}
\ee
In this form, the photon-junction interaction is entirely contained in
the last term.

\subsection{Method of solution}

As in Ref.~\onlinecite{alsaidi}, we diagonalize ${\mathcal{H}}$ in a complete basis
formed from the direct product of the eigenfunctions of
${\mathcal{H}}_{\mathrm{photon}}$ and ${\mathcal{H}}_{\mathrm{J}}^{0}(j)$.
The eigenfunctions of ${\mathcal{H}}_{\mathrm{photon}}$ are, of
course, harmonic oscillator eigenstates; the normalized
$n$th eigenstate has wave function 
$h_n(q)= \frac{1}{\sqrt{\sqrt{\pi}\, 2^{n}\,n! }}
\exp(-y^2/2 )H_{n}(y)$, where $H_n(y)$ is a Hermite polynomial of
order $n$, and $y= (\o/\hb)^{1/2} q$.
An eigenfunction $\psi(\phi_j)$ of ${\mathcal{H}}_{\mathrm{J}}^{0}(j)$
with eigenvalue $E^{(j)}(\nb)$ can be written as
$\psi(\phi_j)=\exp({i\,\bar{n}_j\, \phi_j })\, \eta(\phi_j)$, and
satisfies the Schr\"{o}dinger equation,
${\mathcal{H}}_{\mathrm{J}}^{0}(j)\,\psi(\phi_j)= E^{(j)} \,
\psi(\phi_j)$.  Using the representation $n_j =
-i\,{\partial}/{\partial \phi_j}$, which follows from the commutation
relation $[n_j, \phi_j] = -i$, we write the Schr\"{o}dinger
equation as
\be 
\frac{d^2\,Y(x)}{d\,x^2}+\left(\frac{8\,
E^{(j)}}{U} +2\, Q \cos2\, x \right) Y(x)=0, \label{eq:Mathieu} \ee
where $x = \phi_j/2$, $Y(x)=\eta(\phi_j /2)$, and $Q= 4\, J/U$.  This
is a Mathieu equation with characteristic value $ a^{(j)}={8\,E^{(j)}}/{U}$,
and potential of strength $Q$.  The eigenvalues $E^{(j)}$ are
determined by the requirement  that $\psi(\phi_j)=\psi(\phi_j+2
\pi)$, or equivalently $ Y(x+\pi)=\exp({- 2\, i\, \bar{n}_j\, \pi})\,
Y(x).  $  Thus we are interested only in the
Floquet solutions $Y_{k_j}(x)$ of the
Mathieu equation (\ref{eq:Mathieu}), with Floquet exponent $\nu_j =
2\, k_j -2\, \nb$, where $ k_j=0, \pm 1, \pm 2, \ldots $. We denote
the corresponding eigenvalue of ${\mathcal{H}}_{\mathrm{J}}^{0}(j)$ by
$E_{k_j}^{(j)}(\nb)$.  
It is sufficient to consider only $0
\le \nb \le 0.5 $, because the eigenvalues of
${\mathcal{H}}_{\mathrm{J}}^{0}(j)$ are periodic in $\nb$ with period
unity, and are symmetric about $\nb=1/2$.

Now, any eigenstate $\Psi(\phi_1,...\phi_N, q)$, of the Hamiltonian 
${\mathcal{H}}$ can be expressed as a linear combination of 
product wave functions consisting
of eigenfunctions of ${\mathcal{H}}_{\mathrm{photon}}$ and 
the ${\mathcal{H}}_{\mathrm{J}}^{0}(j)$'s, i.\ ~e.\ ,
\be
\Psi(\phi_1,...\phi_N, q)= \sum_{k_1 \ldots k_N,\,n} 
	A_{k_1\ldots k_N ,\,n}\, h_n(q)\, \prod_{j=1}^{N}\psi_{k_j}(\phi_j) 
\label{eq:psi},
\ee
where $\psi_{k_j}(\phi_j)= \exp(i \,\nb \,\phi_j)
Y_{k_{j}}(\phi_j/2)$, and  $N$ is the number of junctions in the
resonator.  The only term in ${\mathcal{H}}$ which is not diagonal in
this product basis is the interaction term
${\mathcal{H}}_{\mathrm{int}}$.  We can obtain the product coefficients
and eigenvalues by directly diagonalizing the
Hamiltonian of Eq.\ (\ref{eq:hamtotal}) in this product basis, i.\ ~e.,
by solving the Schr\"{o}dinger equation, $ {\mathcal{H}}\,\Psi(\phi,q)=
E\,\Psi(\phi,q),$ where we use the shorthand notation $\phi \equiv
(\phi_1,...\phi_N)$.  Once the eigenvalues $E$ and expansion coefficients
$A_{k_1 \ldots k_N,\,n}$ are known, we can study the 
time evolution of the system  given its state at time $t = 0$,
and hence obtain the time-dependent expectation value of
any operator ${\mathcal{O}}$ at time $t$ using
\be
\langle {\mathcal{O}}(t)\rangle = \mathrm{Tr}\left[\rho(t)\mathcal{O}\right]
\label{eq:timeA}.
\ee
Here $\rho(t)$ is the density matrix of the system at time $t$, defined
by
$
\rho(t)= U(t) \rho(0) U^{\dagger}(t)$, 
where $U(t)= \exp(-i{\mathcal{H}}t/\hb)$
is the evolution operator, and $\rho(0)$ is determined by 
the initial state of the system.

We have carried out extensive calculations for the Hamiltonian (1),
using the product basis described above.  In our calculations we have
arbitrarily used $Q = 0.7$ and $\hbar\,\omega/U = 0.3$, representing a
case where the charging energy, Josephson coupling, and photon energy
are all comparable.  The attainability of these parameters in
realistic systems is discussed further below.  
Most of our
calculations have also been carried out near the value $\bar{n} =
0.258$, corresponding to the case where the energy difference between
the two lowest junction levels equals a single quantum of photon
energy.  In our single-junction calculations, \cite{alsaidi} this
choice leads to resonant absorption of a single photon and to maximum
entanglement between the junction and the photon states.  For larger
number of junctions ( $N \ge 4$) we found that we have to tune
$\bar{n}$ slightly away from the resonance condition of the single
junction to get the system on resonance again.

In all of our calculations, we use a truncated basis of the
eigenstates of ${\mathcal{H}}_{\mathrm{J}}^{0}(j)$, and of
${\mathcal{H}}_{\mathrm{photon}}$.  For $N=1,2,3$ junctions, we
include the first nine photon eigenstates and the first five levels of
each junction, i.\ ~e., a total of $9\, \times\, 5^N$ states.  For
$N=4$, we use fewer eigenstates: the lowest eight photon states and the
first four junction levels.
In all cases, we confirmed the numerical stability of our results by
increasing the dimensionality of the basis and finding that there is
almost no change in our results.  As another check, we have calculated
$\mathrm{Tr}(\rho)$, which should be time-independent and equal to
unity for our normalization.  We find that this trace does remain
equal to unity to high accuracy for all times considered, thus
providing evidence that we have correctly diagonalized the
Hamiltonian.

\section{Results and Discussion}

\subsection{Reduced Hamiltonian}

The main goal of this paper is to study how well the results of this
exact diagonalization procedure can be matched by a reduced Hamiltonian
which includes only the lowest two energy levels from each junction.  For the
one-junction case, the two-level approximation
was already tested in Ref.~\onlinecite{alsaidi} at $\bar{n} = 0.258$.  
At this value of $\bar{n}$, the splitting between the two
lowest levels equals the single-photon energy $\hbar\,\omega$. The reduced 
Hamiltonian then takes the Jaynes-Cummings form: \cite{jaynes}
\be
{\mathcal{H}}_{{JC}} =\hb \, \o \left( a^{\dagger} a + S_{z} \right) + 
 \xi \,a^{\dagger}\, S_{-}+ \xi^{*}\, a \,S_{+}. \label{eq:hamred1}
\ee
Here the junction is represented as a quantum spin-$1/2$, with $[S_{+},
S_{-}]= 2\, S_{z}$, and the junction energy zero is shifted so
that the lowest two levels are $\pm \hbar\,\omega/2$.  The matrix element
coupling the junction and the field is given by $\xi= \la f
|\frac{1}{2}{\mathcal{H}}_{\mathrm{int}}|i\ra $, where
${\mathcal{H}}_{\mathrm{int}}$ is the one-junction version
of expression (\ref{eq:hint}); the ket $|i \ra \equiv | n=1; k=0 \ra $
is a wave function containing one photon and the junction in the ground state
($k=0$), while $|f \ra \equiv | n=0; k=1 \ra $ is a state 
with zero photons and the junction in its first excited state ($k=1$).
To first order in $g$,
${\mathcal{H}}_{\mathrm{int}} \approx - J \, g \sqrt{\o/\hb}\, q \sin \phi
$, and hence, to the same order in $g$, 
\be 
\xi \approx
-({J\,g}/{\sqrt{8}}) \, \la k=0| \sin\phi |k=1 \ra.\label{eq:xi}  
\ee 
This completes the definition of the reduced Hamiltonian for one
junction, and establishes
its relation to the full \h of Eq. (\ref{eq:hamtotal}).

A naive extension of the reduced \h 
(\ref{eq:hamred1}) to $N$ junctions would be
\be
{\mathcal{H}}_{D}= {\mathcal{H}}^{0}_{D}+{\mathcal{H}}^{I}_{D},
\label{eq:hamDicke}
\ee
where 
\bea
{\mathcal{H}}^{0}_{D} &=& \hbar\,\o \,a^{\dagger} a + \hbar\,\o \sum_{j=1}^{N}
					S_{z}^{(j)};\nn\\
{\mathcal{H}}^{I}_{D}&=&  \sum_{j=1}^{N} (\xi\,
   	a^{\dagger} \,S_{-}^{(j)}+ \xi^* \,a\, S_{+}^{(j)}). \nn
\eea
Here $S_{-}^{(j)}$, $S_{+}^{(j)}$, and $S_{z}^{(j)}$ are the
lowering, raising, and inversion operators of the $j$th junction, which
satisfy the commutation relations
\bea
\protect{[}S_{z}^{(j)},S_{\pm}^{(k)}\protect{]} 
		&=& \pm \, S_{\pm}^{(j)} \,\delta_{j k},\nn \\
\protect{[}S_{+}^{(j)},S_{-}^{(k)}\protect{]}  
		&=&  2\, S_{z}^{(j)} \,\delta_{j k}.\nn
\eea
The \h (\ref{eq:hamDicke}) resembles the one-mode
Dicke model, \cite{Dicke,Tav_Cum} which describes $N$ identical two-level atoms
within a radiation wavelength of each other, all interacting with a
single mode resonator.  The initial state problem considered later
in this section has been solved exactly for this Hamiltonian in 
a cavity containing one or more accessible electromagnetic
modes. \cite{dorri,cummings}

For the present system, when the resonator contains more than 
one photon, the Hamiltonian (\ref{eq:hamDicke}) 
does not reproduce the full time-dependent
solution to the Schr\"{o}dinger equation for $N \geq 2$,
at least for our numerical parameters. 
Nevertheless, as shown below, the full numerical solution is still well 
approximated by another reduced Hamiltonian in which
each junction is represented by a basis of two levels only, namely
\be 
{\mathcal{H}}_{D'}=
{\mathcal{H}}^{0}_{D'}+{\mathcal{H}}^{I}_{D'},\label{eq:hamDickeM} 
\ee
where the zeroth order \h ${\mathcal{H}}^{0}_{D'}$ is the same as
the Dicke ${\mathcal{H}}^{0}_{D}$ appearing in Eq.\ (\ref{eq:hamDicke}), but 
the interaction
part of the \h ${\mathcal{H}}^{I}_{D'}$ is now 
\bea
{\mathcal{H}}^{I}_{D'} = \sum_{j=1}^{N} \Big(&\xi& \,a^{\dagger}\,
		S_{-}^{(j)} +  \xi^{*}\, a\, S_{+}^{(j)}\Big)\nn \\ 
		&+&\Omega \sum_{{}^{j,k=1}_{( j\neq k)}}^{N} 
		\left(
			S_{+}^{(j)}S_{-}^{(k)}+S_{-}^{(j)}S_{+}^{(k)} 
		\right).  
\eea
The physics behind this reduced Hamiltonian \cite{comment} is discussed further
below, in Sec. III. D. 2. 

\begin{figure}
\includegraphics[width=9.3cm]{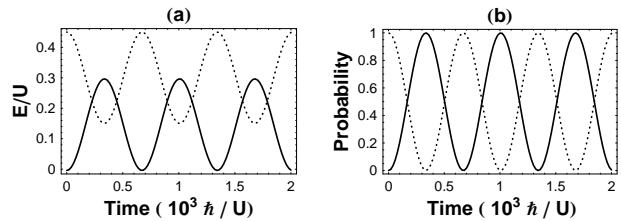}
\caption{\label{fig1}
(a) Time-dependence of $\langle {\mathcal{H}}_{\mathrm{J}}^{0}
\rangle$ (full curve) and of $\langle
{\mathcal{H}}_{\mathrm{photon}}\rangle $ (dashed curve), plotted for
one junction with $\bar{n} = 0.258$ (on-resonance condition). The
system is prepared at time $t = 0$ in the state $| n=1 \,;\,k=0 \ra $.
The other parameters are $Q = 0.7$, $ \omega/U = 0.3$, and $g = 0.15$.
(b) Same as (a) except that we show the time-dependent probability for
finding the Josephson junction (full curve) and the photonic resonator
(dashed curve) in their first excited states.}
\end{figure}

\subsection{Review of single junction}

We first review the time-dependent solution to the Schr\"{o}dinger equation 
for a single junction in a resonator. \cite{alsaidi}  If the 
initial junction-cavity state is $|n=1 \,;\, k=0 \ra$, 
then the system 
evolves in time according to the evolution operator
$U(t)=\exp(-i{\mathcal{H}}\,t/\hb )$.  Fig.\ 1 shows the resulting
time-dependent expectation values $\la {\mathcal{H}}_{\mathrm{J}}^{0}\ra$
and $\la {\mathcal{H}}_{\mathrm{photon}}\ra$, as calculated using Eq.\
(\ref{eq:timeA}) and the full Hamiltonian $\mathcal{H}$.  
Fig.\ 1 also shows the time-dependent probabilities
for finding the junction and the resonator in their first excited
states, given this initial state.  The system wave function is
evidently an entangled state oscillating between $|n=1;\, k =0 \ra$
and $|n=0;\, k =1 \ra$, with a period $T \approx 675 \,\hb /U $
for our particular choice of $g$, $Q/U$ and $ \hbar\,\o /U$.   When the Rabi
period for oscillations between the Josephson-photon entangled states is
calculated using the two-level approximation, it is found to be $\pi\,
\hb/|\xi| = 667\, \hb /U $, in almost perfect agreement with the result
obtained from the full Hamiltonian.

\subsection{Several junctions, asymmetric initial state}

Next, we consider $N$ junctions such that initially (at time $t = 0$)
one junction is in the first excited state, while the remaining $(N -1)
$ junctions are in the ground state, and the cavity initially contains
zero photons. Thus  all the junctions interact with the vacuum fluctuations
of the cavity mode via the Hamiltonian ${\cal H}_{\mathrm{int}}$.

Our primary interest is to compare the time evolution calculated
from the exact Hamiltonian with that obtained from the two reduced
Hamiltonians discussed above.   We will focus our attention on
the ``inversion'' ${\mathcal{S}}(t)$ of the system, defined as
\be
{\mathcal{S}}(t)=  \la S_{z}(t)+\half \ra ,\label{eq:Inv}
\ee 
where the average is taken with respect to the
quantum state  of the system.
${\cal S}(t)$ is the sum of the probabilities that the junctions
are in their first excited states, and satisfies
$0 \leq {\cal S}(t) \leq N$.  This quantity is of interest because
$\hb \, \omega \frac{d{\cal S}}{dt} = I$, is the rate at which energy
is transferred from the junctions to the cavity or vice versa.

\begin{figure}
\includegraphics[width=9.3cm]{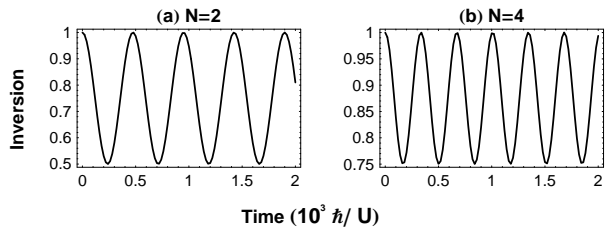}
\caption{\label{fig2}
The time-dependent inversion ${\cal S}(t)$ [Eq.\ (17)] for (a) $N=2$
and (b) $N=4$ junctions in a resonator, calculated using the full
Hamiltonian.  Initially, one junction is in its first excited state,
while all the other junctions and the resonator are in their ground
states.}
\end{figure}

For the initial state considered in this section, it has been
proven\cite{dorri} for the Dicke model [Eq.\ (\ref{eq:hamDicke})] 
that 
\be
{\mathcal{S}}(t) = 1- \frac{1}{N} \sin^2( |\xi|\, \sqrt{N}\, t/ \hb ).
\label{eq:InvN1}
\ee
Thus, the frequency of energy transfer between the photon
field and the junctions is proportional to $\sqrt{N}$.  The group
of $N$ junctions behaves somewhat like a single junction with a coupling to
the cavity mode of $\sqrt{N}\,\xi$ instead of $\xi$.  It also
follows from Eq.\ (\ref{eq:InvN1})
that the photon field never gains more than $1/N$ of the energy initially
stored in the excited junction.  
Thus, as $N$ increases, it becomes progressively less likely that a photon
will be emitted into the cavity, the energy remaining largely trapped within
the junctions ( actually, within the initially excited junction).  
The existence of $N-1$ unexcited
junctions (or ``atoms'') has the effect of preventing the emission of a 
photon from the initially excited junction, \cite{dorri} a phenomenon known as
``radiation suppression.''\cite{cummings}

When we calculate ${\cal S}(t)$ using our direct diagonalization
scheme, we find that the numerical results are in excellent agreement
with the predictions of the two-level approximation (\ref{eq:InvN1})
for $N=1,2,3$ and $4$ junctions in the resonator, at least for several
Rabi periods of oscillation. In Fig.\ 2, we show the inversion as
calculated using the full basis for $N=2$ and $N=4$.

\subsection{Dicke symmetric states}

Next, we consider an initial state in which all the junctions are in
the ground state and there are $n$ photons in the system ($n = 1$,
$2$, $3$).  This is a simple case in which the initial state
of the system is symmetric under the interchange of any two junctions.
We have solved for the time-dependent state of the system using the
full Hamiltonian, and compared the results to the approximate
Hamiltonians (\ref{eq:hamDicke}) and (\ref{eq:hamDickeM}).  Results of
our calculations are shown in Figs.\ 3-5.

As can be seen from these figures, the one-mode Dicke model defined in
 Eq.\ (\ref{eq:hamDicke}) does not
reproduce the full numerical solution for $n \geq 2$.
But the numerical solution is well approximated by the reduced model \h
(\ref{eq:hamDickeM}), in which each junction is still represented by
only two levels.   This Hamiltonian differs from the Dicke model only in
the last term, which is proportional to $\Omega$ and which represents an
effective {\em direct} interaction between the junctions.  
This term is needed in order to
obtain a good fit between the predictions of the reduced model and the
exact Hamiltonian, even though the
junctions in the original model Hamiltonian [Eq. (\ref{eq:hamtotal})] 
have only an indirect coupling with each other through the electromagnetic 
field.

A similar model Hamiltonian to Eq.\ (\ref{eq:hamDickeM}) has been
studied in Ref.~\onlinecite{Joshi} for two atoms in a cavity.  This model
differs from the present one, however, in that the direct coupling in
Ref.~\onlinecite{Joshi} represents a specific term in the Hamiltonian,
namely, an effective dipole-dipole interaction between the atoms.  In
the calculations to be presented below, the coupling strength $\Omega$
in the \h of Eq. (\ref{eq:hamDickeM}) is treated simply as a parameter
determined by best fitting to the numerical data.  However, it can
also be derived explicitly from the full Hamiltonian, as we show
below.

There is one further point to be made about these numerical results.
Namely, so long as $n \le N$, we find that the interaction term $\xi$ which
best fits the computed time evolution is numerically equal to that
which enters the single-junction reduced Hamiltonian [Eq.\
(\ref{eq:xi})] when there is no more than one photon present.  This
result is intuitively reasonable, because in this symmetric case all
the junctions in the resonator share equally in the photons, and thus,
for $n \le  N$, none will absorb more than one photon.
\begin{figure*}
\includegraphics{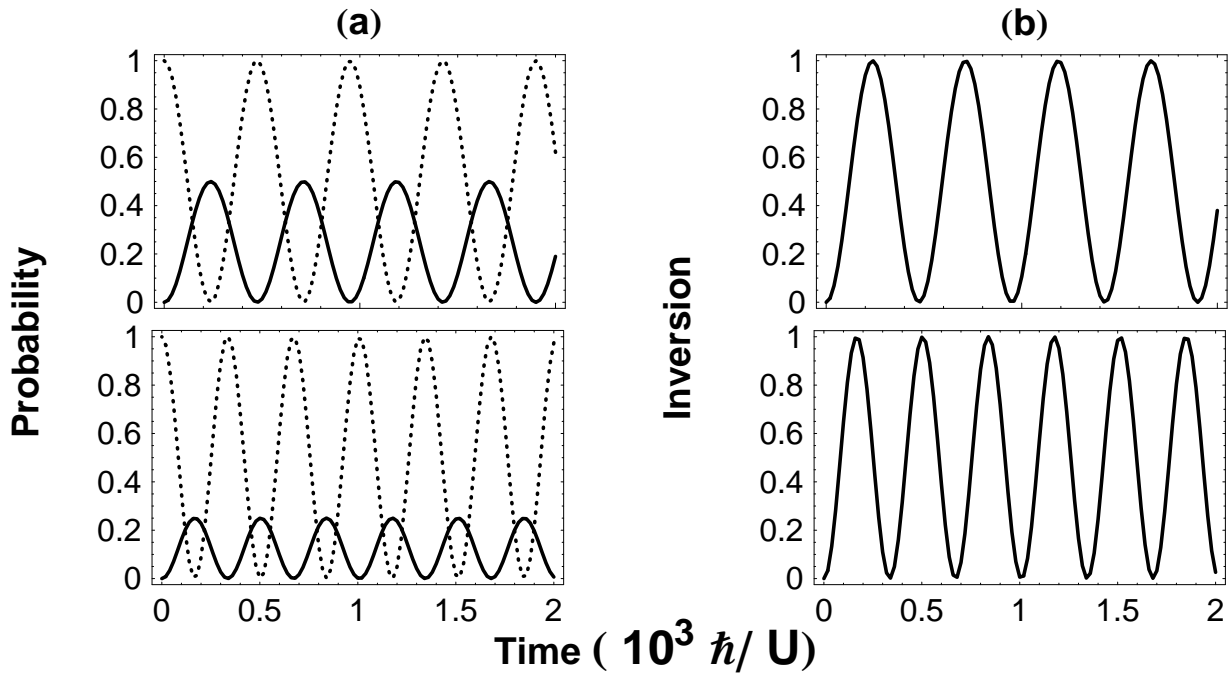}
\caption{\label{fig3}
(a) Time-dependent probability for finding any one of the 
junctions (full curve) or the photonic resonator (dashed curve) in
its first excited state for $N=2$ (top) and $N=4$ (bottom). The system is
prepared at time $t = 0$ such that there is one photon in the system,
and all the junctions are in their ground states. (b) Same as (a)
except we show the inversion ${\cal S}(t)$ of the system as defined in Eq.
(\ref{eq:Inv}).}
\end{figure*}

Now we will discuss the basis used to study the \h  (\ref{eq:hamDickeM}). The
Hilbert space of $N$ two-level junctions is $2^{N}$ dimensional.  Thus
the system is analytically intractable even for a small value of $N$
unless there is some symmetry of the Hamiltonian which confines the
evolution of the system to some lower invariant subspace.
The \h (\ref{eq:hamDickeM}) has two useful symmetries of this kind. 
First, a symmetric initial state is an
eigenstate of the permutation operator ${\mathcal{P}}_{j k} $, which
exchanges the $j$th and $k$th junctions.  Since ${\mathcal{P}}_{j k} $
commutes with the Hamiltonian of Eq. (\ref{eq:hamDickeM}), a symmetric
initial state evolving under ${\cal H}_{D^\prime}$ will remain
symmetric at later times. The second symmetry involves the excitation
number operator
\be
{\mathcal{M}}= 
	a^{\dagger} a + \sum_{j=1}^{N}\left(S_{z}^{(j)}+\half \right) ,
\ee
which also commutes  with
${\mathcal{H}}_{D'}$, and thus is also a constant of the motion. 
In this case it is convenient to study the \h ${\mathcal{H}}_{D'}$ 
using the basis vectors formed from the direct product of the eigenstates of 
${\mathcal M}$ and $a^\dag a$.
We denote these states as $|m; n \ra$, where
\bea
{\mathcal{M}}\,|m; n \ra&=&m \, |m; n \ra, \nn\\
a^{\dagger} a \, |m; n \ra  &=&n \, |m; n \ra,
\eea
and $0 \leq n \leq m$.  Note that the time evolution of the system is
restricted to a $(m+1)$-dimensional subspace spanned by the basis
vectors $|m ; 0\ra, \ldots, |m; m\ra$. \cite{koz}

It is convenient to write out the  state $|m; n \ra $ explicitly.
There are $n $ photon excitations in the cavity, and $m-n$ excitations 
distributed symmetrically among the $N$ junctions.
Thus we may write $| m; n \ra = |n\ra \otimes |m-n \ra_J $, where
$|n\ra $ is a Fock state and 
$|m-n\ra_J$ is a normalized symmetric Dicke state of the junctions,
described by
\be
|k \ra _J=\left[ \frac{N!}{k! ( N- k)!}\right]^{-1/2} 
          \sum_{{\cal{P}}} \prod_{l=1}^{k} |+\ra_{j_l}\prod_{m\neq 1\cdots k} |- \ra_{j_m},
\ee
where $|+\ra_{j_l} \equiv | k_l=1 \ra $ ($|-\ra_{j_l}\equiv | k_l=0
\ra $ ) is the spin up (down) state of the $l$th junction, and the
summation is over permutations ${\cal{P}}$ among the $N$ junctions.

We now discuss the cases of $m=1,2,3$  excitations, and compare
the time-dependent states obtained from the full numerical solution to 
the predictions of the two level approximation, using the basis $|m; n
\ra$.

\begin{figure*}
\includegraphics[width=20cm]{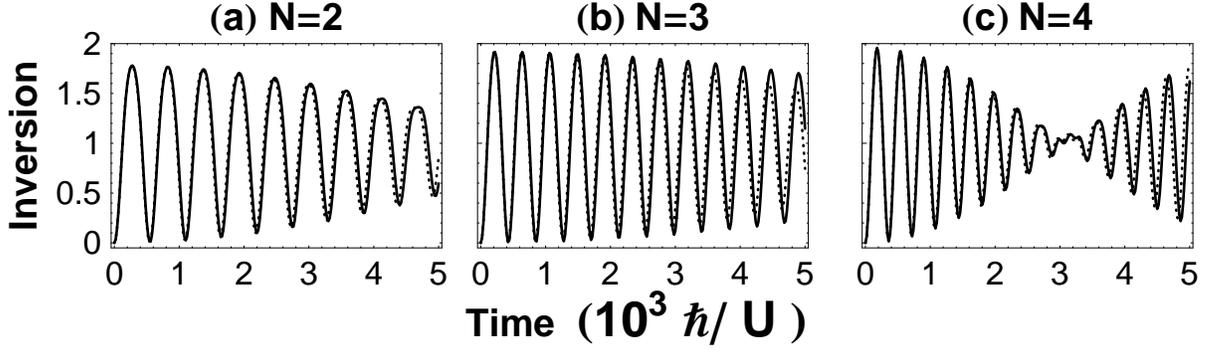}
\caption{\label{fig4}
Time-dependent inversion ${\cal S}(t)$ for $m=2$ excitations and
$N=2,3$ and $4$ junctions in the resonator [(a), (b) and (c)
respectively].  The initial state has two photons in the resonator,
and all the junctions in their ground states. The solid line is the
full numerical solution, and the dotted line follows from
Eqs. (\ref{eq:Inv2S}). Note in particular the decrease in the Rabi
period of oscillation as $N$ increases. Other parameters are the same
as in Fig.\ 1 except that $\nb= 0.2598$ for $N=4$. The Dicke model (~
not shown) would give a sinusoidal oscillation with no beats, both
here and in Fig. \ 5.}
\end{figure*}

\subsubsection{One excitation ($m = 1$)}

In this case, our numerical results are well described by the Dicke \h
[Eq. (\ref{eq:hamDicke})], at least for $N \leq 4$.  This behavior is
expected, since the junctions cannot interact directly with each other
in the presence of only one photon.  For one excitation in the system,
an exact solution for a two-level system can be found even in a damped
cavity. \cite{seke} In the absence of damping, the wave function of the
system in the interaction picture can be written in the two-level
approximation as:
\be
|\Psi^{\mathrm{I}}(t)\ra = C_1(t) |1;1 \ra + C_0(t) |1;0 \ra.
\ee
From the time-dependent Schr\"{o}dinger equation (taking $\hb=1$ henceforth),
\be
{\mathcal{H}}^{\mathrm{I}}_{\mathrm{D}}\,|\Psi^{\mathrm{I}}(t)\ra= i  
\frac{d}{d\,t}|\Psi^{I}(t)\ra, 
\ee
we obtain
\bea
i\, \frac{d C_1(t)}{d\,t}&=& \xi \,\sqrt{N}\,C_0(t), \nn \\
i\, \frac{d C_0(t)}{d\,t}&=& \xi^* \, \sqrt{N}\,C_1(t). \nn
\eea
Taking the junctions initially to be in their ground states and assuming
$n = 1$ photons in the system (i.\ ~e. ~$C_1(t)|_{t=0}=1$), we
find after some algebra that
\bea
 C_1(t) &=&  \cos (|\xi| \, \sqrt{N} \, t ), \nn \\
C_0(t)  &=&  \sin ( |\xi| \, \sqrt{N} \, t  ). \nn 
\eea
Thus the system oscillates between the states $|m=1; n=1 \ra$ and
$|m=1; n=0 \ra$-- that is  the junctions alternately absorb the photon of the
field, then re-emit it, in a reversible evolution, as expected in a
lossless cavity.   The oscillation period $T \propto N^{-1/2}$.
From this solution, the inversion of the system  
[Eq.\ (\ref{eq:Inv})] is 
\be
{\mathcal{S}}(t) = \sin^2( |\xi|\, \sqrt{N}\, t ),  \label{eq:Inv1}
\ee
and hence the rate of radiation into the cavity is
\be
I(t)= - \omega\,\frac{d{\mathcal{S}}(t)}{dt}= - \omega\,|\xi| \sqrt{N} \sin( 2\, |\xi|
\,\sqrt{N}\, t ). \label{eq:power}
\ee
Both the oscillation frequency and the amplitude
of the rate of radiation are proportional to
$\sqrt{N}$, a signature of a cooperative phenomenon. For very short
times, the rate of radiation into the cavity is
\be
I(t) \approx -2 \, \omega \, |\xi|^2 \, N\, t 
\ee
-- that is, $N$ times the rate of a single junction. Thus the
presence of $N$-junctions in the cavity gives rise to a collective
behavior.

Fig.\ 3(a) shows the time-dependent probability that the Josephson
junctions (full curve) and the photonic resonator (dashed curve) are
in their first excited states for $N=2$ and $N=4$.  The system is
prepared at time $t=0$ with all the junctions in the ground state, and
the resonator in the number state $|n=1\ra$.  The system develops into
an entangled state such that the photon is shared equally among all
the junctions.  This behavior can be seen both in the two-level
approximation and in the results of an exact calculation.  In Fig.\
3(b) we show the inversion, ${\mathcal{S}}(t)$, calculated using the
full Hamiltonian for $N = 2$ and $N=4$ junctions and one excitation ($m
= 1$).  The numerical results are clearly in excellent agreement with
the predictions of the two-level approximation [Eq. (\ref{eq:Inv1})].

\subsubsection{Two excitations ($m = 2$)}

For $m=2$ we need consider only three eigenstates of the Hamiltonian.
In the interaction picture the system wave function is
\be
|\Psi^{\mathrm{I}}(t) \ra = C_2(t)|2 ; 2 \ra +  C_1(t)|2 ; 1 \ra +
C_0(t)|2 ; 0 \ra, \label{eq:psi2} 
\ee
where the coefficients satisfy 
\bea
i \, \frac{d C_2(t)}{d\,t} &=& \xi \,\sqrt{2 \,N } \,C_1(t) ,\nn \\
i \, \frac{d C_1(t)}{d\,t} &=&  \xi^* \,\sqrt{2 \,N } \,C_2(t) +
		     \Omega \,C_1(t) + \xi\, \sqrt{2\,(N-1)
						}\,C_0(t),\nn \\	
i \, \frac{d C_0(t)}{d\,t} &=& \xi^* \,\sqrt{2\,(N-1) }\, C_1(t). \nn	
\eea
Assuming an initial state $|2 ; 2\ra$, i. e. a state containing 
two photons, with all the junctions in their ground states, we find
that the solution of these coupled equations is
\begin{widetext}
\bea
C_2(t) &=&\frac{1}{\zeta\,(2\,N-1)} 
		\left\{ 
		    \zeta \,(N-1)+N\,e^{- i\, \Omega\, t /2}
		         \left[ 
			 \zeta \cos(\zeta \,t) + i\, \Omega \sin(\zeta \, t)
			\right] 
		 \right\}, \nn \\
C_1(t) &=& - i\,\sqrt{2\,N}\left(\xi /  \, \zeta\right)\,
	e^{-\half \, i\,\Omega\,t} \sin (\zeta \,t),
		  \label{eq:Inv2S} \\	
C_0(t) &=&\frac{\sqrt{N\,(N-1)}}{2 \,(2\, N-1) \zeta}\, 
		\left\{ e^{-\half\, i\,\Omega \,t } \,
			\left[2\, \zeta \,\cos(\zeta\, t)
			+ i \,\Omega \,\sin(\zeta\, t)
			\right]-2\,\zeta 
		\right\},\nn
\eea
where $\zeta= \half \, \sqrt{\Omega^2+ 8\, |\xi|^2 ( 2\,N -1)}$. 
The probability of a two-junction excitation of the system is $P_2(t)= |
C_0(t)|^2 $, which can be reduced to
\bea
P_2(t)= \frac{N(N-1)}{(2\,N-1)^2 \zeta^2}\, 
	\Big[ 2\,\zeta^2&-&2 \, (2 \, N-1)\,|\xi|^2 \,\sin^2 (\zeta\,t )  \nn \\
	&-&
  2\,\zeta^2 \cos (\zeta\,t )\,\cos (\Omega \,t/2 )   - 
     \Omega \,\zeta \,\sin (\zeta\,t)\,\sin (\Omega \, t/2)
	\Big].
\eea
\end{widetext}
For $\Omega=0$, this becomes
\be
P_2(t)= \frac{N(N-1)}{(2\,N-1)^2 }\, 
	\left[ \cos(\omega_2 t)-1\right]^2,
\ee
so that the Rabi period of oscillation is now $\o_2=[ 2
(2\,N-1)]^{1/2}\, |\xi| $.   Since $|P_2(t)| \leq N(N-1)/(2N-1)^2 \leq 1/4$,
the system is unlikely to contain two excitations simultaneously.  Even
if $\Omega \neq 0$, the same statement is still valid, but the time evolution
is no longer perfectly sinusoidal, but instead is characterized by
 beats with a period inversely proportional to $\Omega$.

The inversion of the system is given by
\be
{\mathcal{S}}(t)= | C_1(t)|^2+2\,| C_0(t)|^2.\label{eq:Inv2}
\ee
A full expression for the inversion readily calculated from Eqs.\
(\ref{eq:Inv2S}) is quite complicated, but is not given here.  Fig. 4
shows the results obtained numerically for $N = 2, 3$, and $4$,
using the full basis (solid line), as well as the inversion calculated
from Eq. (\ref{eq:Inv2}) using the modified Dicke model with $\Omega
\neq 0$ (dashed line).

We now discuss the origin of the extra direct interaction in the
effective Hamiltonian (\ref{eq:hamDickeM}), in light of these
numerical results.  Eq.\ (\ref{eq:psi2}) expresses the wave function
for two excitations as a linear combination of three states, which we
will denote compactly as $|a\ra $, $|b\ra $, and $|c\ra $
respectively.  In the absence of ${\cal H}_{\mathrm{int}}$, these
states are degenerate.  According to the Dicke model, ${\cal
H}_{\mathrm{int}}$ breaks this degeneracy because some of the matrix
elements of ${\cal H}_{\mathrm{int}}$ between these states are
nonzero.  This is the {\em first-order} effect of ${\cal
H}_{\mathrm{int}}$ (or, equivalently, in $g$).

However, there is also a {\em second-order} effect.  Namely,
as is well known from degenerate second-order L\"owdin perturbation 
theory, \cite{lowdin} the {\em effective} Hamiltonian  matrix elements between,
for example, states $\la a|$ and $|b\ra $ should be written
\begin{equation}
\la a|{\cal H}_{\mathrm {eff}}|b\ra  =
\la a|{\cal H}_{\mathrm{int}}|b\ra  + \sum_r^\prime\frac{\la a|{\cal H}_{\mathrm{int}}|r\ra 
\la r|{\cal H}_{\mathrm{int}}|b\ra }{E_a - E_r}.
\end{equation}
Here the sum runs over all states $|r\ra $ {\em other than} 
$|a\ra $, $|b\ra $, and $|c\ra $.  These are the terms which give rise
to the extra interaction proportional to $\Omega$ in Eq. (\ref{eq:hamDickeM}).

To confirm this hypothesis, we have calculated these matrix
elements, including the terms second order in ${\cal H}_{\mathrm{int}}$,
through order $g^2$.   We do find that these terms produce
a direct interaction between the junctions of the form 
(\ref{eq:hamDickeM}), as found numerically.  Specifically,
the matrix element $\la b|{\cal H}_{\mathrm{eff}}|b\ra $ becomes nonzero when
these second-order terms are included.  Furthermore, we find 
that the value of $\Omega$ found from this direct calculation
is comparable to that found numerically.  Specifically, for
the the cases mentioned above, we find $\Omega/(|\xi|^2/\omega)
=4.89$, $9.65$, $14.42$  for $N = 2$, $3$, and $4$. For
comparison, the best fit values of $\Omega$  to the numerical
results of Fig.~4 are $\Omega/ (| \xi|^2/\omega) \approx
5.74, 4.46, 13.40 $  for $N=2,3$ and $4$ respectively.

\subsubsection{Three excitations ($m = 3$)}

In this case the wave function at any time $t$ in the interaction
picture is
\be
|\Psi^{\mathrm{I}} (t)\ra= \sum _{i=0}^{3}\,C_i(t)\,|3 ; i \ra ,
\ee
where  the expansion coefficients satisfy
\begin{widetext}
\bea
i \, \frac{d C_3(t)}{d\,t} &=& \xi \,\sqrt{3 \,N } \,C_2(t), \nn \\
i \, \frac{d C_2(t)}{d\,t} &=&  \xi^* \,\sqrt{3 \,N } \,C_3(t) +
		     \Omega \,C_2(t) + 2\,\xi\, \sqrt{N-1 }\,C_1(t),\nn\\	
i \, \frac{d C_1(t)}{d\,t} &=& 2\,\xi^* \,\sqrt{N-1 }\, C_2(t) 
		+ \Omega \,C_1(t) + \xi \, \sqrt{3\, ( N-2)}\,
				C_0(t), \nn \\
i \, \frac{d C_0(t)}{d\,t} &=&   \xi^* \, \sqrt{3\, ( N-2)} \,C_1(t). \nn
\eea
\end{widetext}
The inversion of the system in this case is
\be
{\mathcal{S}}(t)= | C_2(t)|^2+ 2\,| C_1(t)|^2+ 3\,| C_0(t)|^2.\label{eq:Inv3}
\ee
In Fig. 5 we show the results obtained from the full numerical
solution (full line) and from Eq. (\ref{eq:Inv3}) (dashed line) for
$N=3,$ and $N=4$ junctions.  
Here the best fit values of $\Omega$ are given such that  
$\Omega/ (|\xi|^2/\o) \approx
7$ and $13.4$  respectively. For comparison, the second-order
degenerate perturbation calculation mentioned above gives
$\Omega/(|\xi|^2/\o)= 10$ and $15.5$.

\begin{figure}
\includegraphics[width=10cm]{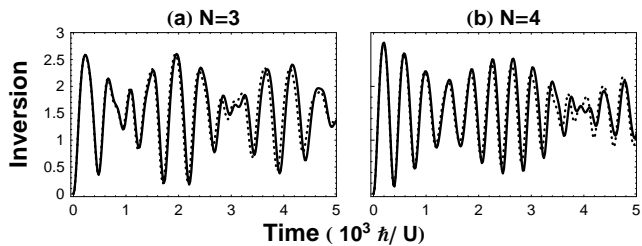}
\caption{\label{fig5}
Same as Fig.\ 3, except that the initial state has three photons in
the resonator, and calculations are carried out for $N=3$ and $N=4$ junctions,
as indicated in the figure.}
\end{figure}

\subsection{Coherent state}

As a last non-trivial case we consider the system such the resonator
is initially in a coherent state. This case is of special interest
because the number eigenstates of the electromagnetic field are
generally difficult to generate. Nevertheless, as we will see below,
our main conclusion, namely the enhancement of the junction-resonator
interaction with increasing $N$ still holds in this case.

A coherent state $|\beta \ra $ with an average number of photons $\la n
\ra = | \beta |^2 $  can be written as a linear superposition of  
number states $|s \ra$ with a Poisson distribution:  
\be
|\beta \ra = \sum_{s} e^{- |\beta|^2 /2}\, \frac{\beta^s}{\sqrt{s!}}\, | s
\ra.
\ee
For $N=1$, it is  easily
proved that the inversion of the system is given by:
\be
{\mathcal{S}}(t) = \frac{1}{2}\, - \frac{1}{2} \, e^{- |\beta|^2 }\,\sum_{s=0}^{\infty}\,\frac{ |\beta|^{2 s}}{s!} \, \cos( 2\, \xi
\,\sqrt{s}\,t),
\ee
which shows the phenomenon of `` collapse and revival'' i.\ e. the
envelope of the Rabi oscillations periodically ``collapses'' to a half
with a subsequent ``revival'' at later times. \cite{alsaidi,eberly} 
For $N > 1$, the analysis of the previous
subsection can be readily generalized to treat a coherent initial
state.

As an example, we show in Fig.\ 6 the inversion ${\mathcal{S}}(t)$ for
$N=1$ and $N=2$  assuming  $\beta =
\sqrt{2}$. The solid lines follow from the full Hamiltonian while the
dashed lines are from the reduced Hamiltonians of
Eqs. (\ref{eq:hamred1}) and (\ref{eq:hamDickeM}). The best fit value
of $\Omega $ in this case is $\Omega \approx 5.74 \, |\xi|^2/\o $. 
This value of $\Omega$ is
the same as for the number state considered before with $m=2$
excitations. The enhancement of the junction-cavity interaction can be
seen from the decrease of the period of oscillation in Fig. \ 6(b)
compared to Fig. \ 6(a).

\begin{figure}
\includegraphics[width=10cm]{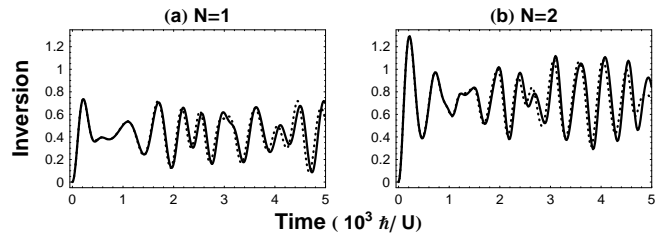}
\caption{\label{fig6}
The time-dependent inversion ${\cal S}(t)$ for (a) $N=1$ and (b) $N=2$
junctions given that the initial state of the resonator is a coherent
state with average photon number $\la n \ra =2 $, and the junctions
are initially in their ground states. The solid line denotes the full
numerical solution, and the dotted line follows from the two-level
approximation.}
\end{figure}

\section{Summary and Conclusions}

The present paper extends our previous publication\cite{alsaidi} to
the case of $N >1$ identical junctions interacting with a single
lossless cavity mode.  Although oversimplified, this model has the
advantage that we can obtain detailed, numerically exact dynamical
solutions with a range of initial conditions.  Our treatment of the \h
is entirely quantum-mechanical.

We have diagonalized the \h exactly, using a basis large enough to
insure accuracy, and compared our results to those of a reduced
Hamiltonian [Eqs. (\ref{eq:hamDicke}) or (\ref{eq:hamDickeM}) ], in
which each junction is represented by two levels only. The reduced \h
resembles the one-mode Dicke \h but contains a direct interaction term
between the junctions.  For short times (of the order of several Rabi
periods) the agreement between results obtained from the two-level \h
of Eqs.  (\ref{eq:hamDicke}) or (\ref{eq:hamDickeM}) and those found
from the full \h [Eq. (\ref{eq:hamtotal})] is excellent, but decreases
somewhat for longer times.  Also, as expected for a Dicke-like model,
the system of several identical junctions behaves just like a single
junction for short times, but with an enhanced junction-cavity
coupling strength $\xi \, \sqrt{N}$, and a correspondingly decreased
Rabi period. This reduced Rabi period might be smaller than the
coherence time, making these oscillations experimentally observable.

The most striking feature of our numerical results is the presence of
an effective {\em direct} interaction between the junctions resulting
from the cavity.  This interaction has the form of a dipole-dipole
interaction between the ``spins'' representing the two lowest levels
of each junction.  This extra interaction must be included in the
effective two-level Hamiltonian in order for it to reproduce the
numerical results obtained from the full Hamiltonian at long times.
While the analogous term in atomic physics is a true dipole-dipole
interaction, in our model it arises through the effects of the higher
energy levels on the degenerates states, as calculated through second
order L\"owdin degenerate perturbation theory.

Before concluding, we briefly discuss the case of non-identical
junctions.  Obviously, this is an important consideration for
Josephson junctions since, in contrast to atoms, it is impossible
to fabricate perfectly identical Josephson junctions.
To test whether the coherence survives when the junctions are
non-identical, we have repeated the two-junction calculations
shown in Figs.\ 2-4 using junctions which differ slightly from
one another.  For convenience, we considered two junctions with
slightly different values of $J$, but the same values of $U$.
With these parameters, the resonant values of $\bar{n}$ are
slightly different for the two junctions in the same cavity.
We find that, if the two $J$'s differ only slightly (by only
1-3\% for our choice of $J/U$), the coherent behavior, as signaled
by the $\sqrt{2}$ enhancement of the Rabi oscillation frequency, 
survives.  For greater differences between the $J$'s,
the coherence disappears and one finds instead two distinct
resonances, each with an unenhanced Rabi oscillation frequency.  

We have also carried out similar calculations in other parameter
regimes - specifically, for $Q$ varying between $0.01$ and
about $3$ (with $g$ remaining unchanged).  The qualitative
results - that is, the presence of Rabi oscillations and the
characteristic frequency changes associated with coherence -
are unchanged.

Finally, we briefly discuss whether the present system could be
constructed in the laboratory.  The values of the parameters entering
the model have been estimated previously.  \cite{alsaidi} For a
cylindrical cavity of length $d$ and radius $r$, a typical $g$ [ Eq.
(\ref{eq:g})] was found to be $g \sim 3\,\ell\,\sqrt{\alpha/(rd)}$,
where $\alpha = e^2/(\hbar c) \sim 1/137$ is the fine structure
constant, and $\ell_\|$ is the junction thickness.  Achievable values
of the other parameters were estimated as $U \sim 10^{-15}$ erg $\sim
8 K$, and $\omega \sim 2c/r$.  In order to attain $Q \sim 0.7$, as
used in our calculations, we need $J/k_B \sim 3$ K, or a critical
current $I_c \sim 0.1 \, \mu$A.  To obtain $\hbar\,\omega/U \sim 0.3$,
these estimates give $x \sim 10^{-3}$ cm and $r \sim 0.1 - 1$
cm, where $x$ is a typical junction dimension (e.\ ~g., island size,
junction radius).  These values, which would give $\omega \sim 100$
GHz, are achievable in present technology. 
The corresponding maximum power radiated into the
cavity [Eq. \ref{eq:power}] will be of the order of $10^{-11} \sqrt{N}$~W. 

However, the value $g = 0.15$, used in our calculations, is much
larger than the value $g \sim 10^{-5}$ which was previously
estimated\cite{alsaidi} to result from the parameters just described.
Such a small $g$ would certainly cause the two-level approximation to
be even better than found in our calculations here.  But it would have
the disadvantage of producing longer Rabi oscillation time, and hence,
of making coherence over a Rabi period more difficult to achieve;
it would also lead to a much lower power radiated.  Thus, the
largest possible $g$ is clearly desirable in order to detect this
effect.  The Rabi period could also be 
decreased by junctions with larger critical currents than those
considered here.  However, the critical current is 
limited by the requirement $\hb\,
\omega \ll \Delta$, where $\Delta$ is the superconducting gap.
Thus, the best hope for observing these oscillations may be
to reduce the decoherence effects.

\begin{acknowledgments} 
We thank professors G.~P.~Lafyatis and D.~W.~Schumacher for useful
conversations. This work has been supported by the National Science
Foundation, through Grants DMR97-31511 and DMR01-04987.  
Computational support was provided, in part, by the Ohio 
Supercomputer Center.
\end{acknowledgments}


\begin{thebibliography}{3}
\bibitem{comment}In Ref.~\onlinecite{MSS},
a direct coupling between the qubits was obtained by tracing out the photon
variables, a calculation which was feasible because the quantum of radiation
was large compared to the Josephson energies $U$ and $J$. 
In the present calculation, by contrast, we have only a few photons
whose energy is comparable to $U$ and $J$.
\end{thebibliography}
\end{document}